\documentclass[conference]{IEEEtran}
\IEEEoverridecommandlockouts
\usepackage{cite}
\usepackage{booktabs}
\usepackage{hyperref}
\usepackage{amsmath,amssymb,amsfonts}
\usepackage{algorithmic}
\usepackage{graphicx}
\usepackage{textcomp}
\def\BibTeX{{\rm B\kern-.05em{\sc i\kern-.025em b}\kern-.08em
    T\kern-.1667em\lower.7ex\hbox{E}\kern-.125emX}}
\begin{document}

\title{Vintage Code, Modern Judges: Meta-Validation in Low Data Regimes}

% \author{
% \IEEEauthorblockN{Anonymous Authors}
% }

% \author{
% Ora Fandina \quad Gal Amram \quad Eitan Farchi \quad Shmulik Froimovich \\
% Raviv Gal \quad Wesam Ibraheem \quad Rami Katan \quad Alice Podolsky \quad Orna Raz \\
% IBM Research, Israel \\
% \texttt{\{ora.nova.fandina, gal.amram, farchi, shmulik.froimovich, ravivg, wesam, rami.katan, alice.podolsky, ornar\}@ibm.com}
% }

% \author{
% \IEEEauthorblockN{Ora Fandina}
% \IEEEauthorblockA{
% \textit{IBM Research, Israel} \\
% %ora.nova.fandina@ibm.com}
% }
% \and
% \IEEEauthorblockN{ Gal Amram}
% \IEEEauthorblockA{
% \textit{IBM Research, Israel} \\
% %gal.amram@ibm.com}
% }
% \and
% \IEEEauthorblockN{Eitan Farchi}
% \IEEEauthorblockA{
% \textit{IBM Research, Israel} \\
% FARCHI@il.ibm.com}
% \and
% \IEEEauthorblockN{Shmulik Froimovich}
% \IEEEauthorblockA{
% \textit{IBM Research, Israel} \\
% Shmulik.Froimovich@ibm.com}
% \and
% \IEEEauthorblockN{Raviv Gal}
% \IEEEauthorblockA{
% \textit{IBM Research, Israel} \\
% RAVIVG@il.ibm.com}

% \and
% \IEEEauthorblockN{Wesam Ibraheem}
% \IEEEauthorblockA{
% \textit{IBM Research, Israel} \\
% WESAM@il.ibm.com}
% \and
% \IEEEauthorblockN{Rami Katan}
% \IEEEauthorblockA{
% \textit{IBM Research, Israel} \\
% Rami.Katan@il.ibm.com}

% \and
% \IEEEauthorblockN{Alice Podolsky}
% \IEEEauthorblockA{
% \textit{IBM Research, Israel} \\
% Alice.Podolsky@ibm.com}
% \and
% \IEEEauthorblockN{Orna Raz}
% \IEEEauthorblockA{
% \textit{IBM Research, Israel} \\
% ORNAR@il.ibm.com}
% }

\author{
\IEEEauthorblockN{
Ora Fandina, Gal Amram, Eitan Farchi, Shmulik Froimovich, Raviv Gal \\
Wesam Ibraheem, Rami Katan, Alice Podolsky, and Orna Raz}
\IEEEauthorblockA{
IBM Research, Israel \\
%\texttt{
\{ora.nova.fandina, gal.amram, farchi, shmulik.froimovich, ravivg, wesam,\\
rami.katan, alice.podolsky, ornar\}@ibm.com}
%}
}

\maketitle

\begin{abstract}
Application modernization in legacy languages such as COBOL, PL/I, and REXX faces an acute shortage of resources, both in expert availability and in high-quality human evaluation data. While Large Language Models as a Judge (LaaJ) offer a scalable alternative to expert review, their reliability must be validated before being trusted in high-stakes workflows. Without principled validation, organizations risk a circular evaluation loop, where unverified LaaJs are used to assess model outputs, potentially reinforcing unreliable judgments and compromising downstream deployment decisions.

Although various automated approaches to validating LaaJs have been proposed, alignment with human judgment remains a widely used and conceptually grounded validation strategy. In many real-world domains, the availability of human-labeled evaluation data is severely limited, making it difficult to assess how well a LaaJ aligns with human judgment.

We introduce SparseAlign, a formal framework for assessing LaaJ alignment with sparse human-labeled data. SparseAlign combines a novel pairwise-confidence concept with a score-sensitive alignment metric that jointly capture ranking consistency and score proximity, enabling reliable evaluator selection even when traditional statistical methods are ineffective due to limited annotated examples.

SparseAlign was applied internally to select LaaJs for COBOL code explanation. The top-aligned evaluators were integrated into assessment workflows, guiding model release decisions.

We present a case study of four LaaJs to demonstrate SparseAlign’s utility in real-world evaluation scenarios.
\end{abstract}

\section{Introduction}

%\subsection{Legacy Code Modernization: Scarcity of Data and Expertise}
The modernization of legacy software systems, particularly those implemented in languages such as COBOL, PL/I, and REXX, remains a strategic imperative for organizations operating critical infrastructure. However, this process is constrained by severe resource limitations, most notably the diminishing availability of domain experts and the absence of large, high-quality datasets for evaluation purposes. As noted in recent work on assessing code transformations in modernization pipelines, the lack of gold-standard references and task-specific benchmarks significantly hinders the reliable evaluation of system outputs~\cite{froimovich2025qualityevaluationcoboljava}

%\subsection{Large Language Models as Judges: Opportunities and Limitations}
In response to the scarcity of expert evaluators, {Large Language Models as a Judge} have emerged as a scalable alternative to human assessment for tasks such as code translation, refactoring validation, and documentation generation. These systems leverage the generative and reasoning capabilities of large-language models to provide flexible, semantic evaluations beyond surface-level similarity~\cite{li2024llmsasjudgescomprehensivesurveyllmbased}. However, LaaJs exhibit well-documented weaknesses, including susceptibility to positional and verbosity biases, prompt sensitivity, and self-preference effects~\cite{ye2024justiceprejudicequantifyingbiases}. Such vulnerabilities raise concerns regarding their suitability for deployment in high-stakes workflows, where subtle inaccuracies in judgment may lead to costly or dangerous outcomes.

\begin{figure}[h]
    \centering
    \includegraphics[width=0.9\linewidth]{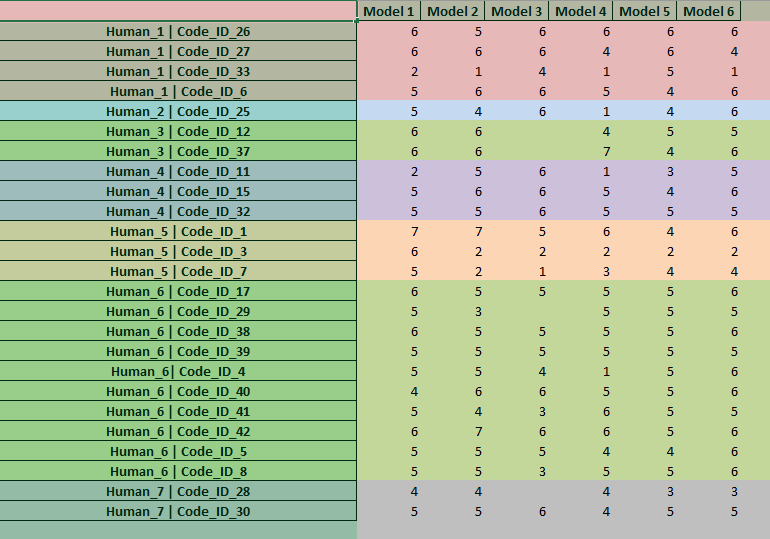}
    \caption{Human evaluation on COBOL: Six models explained the same 25 samples; seven annotators received sparse, disjoint subsets.}
    \label{fig:human_eval_data}
\end{figure}

Given these vulnerabilities, rigorous validation of LaaJs is essential prior to their adoption in production workflows. Several methodologies have been proposed for evaluator validation~\cite{10638599, fandina2025automatedvalidationllmbasedevaluators}. 

A commonly adopted strategy is to measure alignment with human evaluations, treating human-derived rankings as the reference standard. However, this approach presupposes access to sufficient and well-distributed human annotations, which is often violated in practice, particularly in specialized domains where expert evaluations are scarce and costly. For example, in one instance of our real-world internal settings, we had as few as 25 annotated samples per model (see Figure~\ref{fig:human_eval_data}). Under such constraints, conventional statistical tools may yield inconclusive or unstable results.

To address this gap, we introduce { SparseAlign}, a lightweight, interpretable framework designed to validate evaluator alignment under extreme data sparsity.\\
\noindent
{\bf Alignment Under Sparse Human Data}. 
{SparseAlign} provides a principled formalism for quantifying agreement between LLM-based evaluators and expert human judgments in low-data regimes. Specifically designed for industrial settings where annotated examples are scarce, it builds on and extends techniques from order statistics.

The key contributions are as follows: 
1) {\it Sparse Signal Extraction}: A method for inferring model ranking signals from non-overlapping, minimal human feedback. 2) {\it Confidence-Weighted Comparison}: A novel definition of pairwise confidence to quantify the reliability of individual ranking preferences.
3) {\it Alignment Metric}: An interpretable alignment score that combines confidence-weighted ranking agreement and absolute score proximity, yielding a robust measure of evaluator reliability.

While our framework is broadly applicable across tasks and evaluation metrics, we demonstrate its utility on a use case drawn from internal modernization workflows: ranking LaaJs by their ability to generate accurate explanations for COBOL programs. \\
\noindent
{\bf Related Work}.
 Recent surveys outline LaaJ applications and limitations ~\cite{li2024llmsasjudgescomprehensivesurveyllmbased, gu2025surveyllmasajudge}. Validating these evaluators remains difficult under sparse supervision: Guerdan \emph{et al.} show that standard protocols may favor weak judges in low-agreement settings~\cite{guerdan2025validatingllmasajudgesystemsabsence}, while Li \emph{et al.} propose rubric-based multi-agent setups that assume dense annotations~\cite{li2025leveragingllmsmetajudgesmultiagent}.

Legacy code modernization poses further challenges due to limited benchmarks. Diggs \emph{et al.} show that automated metrics poorly reflect human preferences in legacy systems~\cite{diggs2024leveragingllmslegacycode}. Vo \emph{et al.} find LaaJs effective for Bash code evaluation when enhanced with bidirectional matching~\cite{vo2025llmasajudgereferencelessautomaticcode}.

\noindent
{\it Ongoing Work}. To validate SparseAlign’s robustness, we adopt a simulation-based approach~\cite{amram2025laajmeterframeworklaajevaluation}, using virtual LaaJs with tunable quality to test metric sensitivity under controlled noise. While the full framework is under development, we present a preliminary variant with two virtual LaaJs to illustrate SparseAlign’s ability to distinguish evaluator quality.

\section{Details and Experiments}
% begin by describing the precise setup used within our framework.

%\subsection{Framework Setup and Definitions} 
Let ${M_1, M_2, \ldots, M_k}$ denote LLM models, each of which generates explanations for a predefined set of COBOL code snippets. Let $(C_i, E_i)$ represent the input–output pair for model $M_i$, where $C_i$ is the set of COBOL snippets and $E_i$ is the corresponding set of explanations generated by $M_i$.

The outputs are evaluated by a set of human experts ${H_1, \ldots, H_t}$ and a set of candidate LaaJ evaluators ${\text{LaaJ}_1, \ldots, \text{LaaJ}_n}$. Each human expert $H_j$ is assigned a subset of the samples from $C_i$ for evaluation. We denote this subset as $(C_i[H_j], E_i[H_j]) \subseteq (C_i, E_i)$.

%representing the COBOL samples and their explanations from model $M_i$ that were evaluated by human $H_j$.

We assume limited evaluation data, with the total number of samples \(\sum_i |C_i|\) being small. SparseAlign is designed to operate even under extreme human evaluation data constraints, such as when each sample is rated by only one evaluator (i.e., \(C_i[H_j] \cap C_i[H_{j'}] = \emptyset\) for \(j \ne j'\)), which precludes traditional inter-annotator agreement analysis and necessitates alternative strategies for aggregating and comparing model-level rankings.

Each human evaluator scores their assigned samples on a fixed scale from \(1\) to \(s\), using shared evaluation guidelines to promote consistency. Each candidate LaaJ also scores { \it all samples} on the same scale, but operates according to its own instruction prompt, which defines its evaluation behavior and may vary significantly across  candidates and serve as the primary configuration parameter for each LaaJ.

The framework consists of two main parts: 

\begin{enumerate}

\item {\bf Human-Rank}: An algorithm for deriving a reliable consensus ranking of models from limited, disjoint human evaluation data.

\item {\bf Align-Score}: A metric that quantifies how closely a candidate LaaJ replicates the human-derived model ranking.
\end{enumerate}

\subsection{Human-Rank Algorithm}
To derive a consensus model ranking from sparse and imbalanced human evaluations, we begin by computing the average score for each model across all samples it was evaluated on. 

When the difference between the average scores of two models, say $M_i$ and $M_j$, is smaller than a threshold $\delta = \frac{1}{6} \cdot \text{median}\{\sigma_i\}_{i=1}^{k}$, where $\sigma_i$ is the standard deviation of scores for model $M_i$, we consider the models tied in the global ranking. This threshold scales with the typical variability in human scores across models, ensuring that only differences substantially larger than the inherent score dispersion are considered meaningful.

A limitation of this rule is that the tie relation is not transitive: it is possible for 
$M_1$ to tie with $M_2$,  and for $M_2$ to tie with $M_3$, while $M_1$ and $M_3$ differ by more than $\delta$.

To reduce such inconsistencies, we apply a greedy clustering procedure: models are sorted by their average score in decreasing order, and ties are formed sequentially, adding a model to the current tie group only if its inclusion does not cause the group’s score span to exceed 
$\delta$ threshold.\\
\noindent
\textbf{Greedy Tie-Clustering.}
Set $
\delta = \frac{1}{6} \cdot \mathrm{median}\{\sigma_i\}_{i=1}^k
$
where $\sigma_i$ is the standard deviation of scores for model $M_i$.
Sort models by average score $\mu_i$ in descending order (highest to lowest). \\
Start the first cluster with the top model. \\
For each remaining model $M_i$ in sorted order:
\begin{enumerate}
    \item Check if its $\mu_i$ differs by at most $\delta$ from every model in the current (last) cluster.
    \item If yes, add it to that cluster.
    \item If no, start a new cluster with this model.
\end{enumerate}

Output the clusters in the order they were formed.

This procedure produces a partial order over the models, where each model belongs to exactly one tie cluster. A cluster may contain a single model or multiple models whose relative preference is interpreted as undecidable. \\

\noindent
{\bf Breaking Ties: Weighted Pairwise Voting with Confidence}. When models are placed in the same tie cluster because their average human scores differ by less than a threshold $\delta$, we refine the ordering within the cluster using \emph{weighted pairwise voting} with an associated \emph{confidence score}. The goal is to use the strongest available human preferences to refine ties, while ensuring that no preference cycles are created and that ties remain when evidence is weak or conflicting.

Let $\mathcal{H}_{i,j}$ be the set of humans who evaluated outputs from both $M_i$ and $M_j$. For each human $H \in \mathcal{H}_{i,j}$:
\begin{itemize}
    \item Let $\mu_H^i$ and $\mu_H^j$ be the average scores $H$ assigned to outputs from $M_i$ and $M_j$, respectively.
    \item Let $n_H = |S_H^i| + |S_H^j|$ be the total number of samples $H$ evaluated for this pair.
    \item Let $w_H = \frac{n_H}{\sum_{H' \in \mathcal{H}_{i,j}} n_{H'}}$ be the normalized weight of annotator $H$ for this pair.
\end{itemize}

Each annotator contributes $w_H$ to the model with the higher average score for this pair; in case of equality, $w_H$ is split equally. The total weighted votes are:
\[
V(M_i, M_j) = \sum_{H \in \mathcal{H}_{i,j}} w_H \cdot 1[\mu_H^i > \mu_H^j] + \tfrac{1}{2} w_H \cdot 1[\mu_H^i = \mu_H^j]
\]
\[
V(M_j, M_i) = \sum_{H \in \mathcal{H}_{i,j}} w_H \cdot 1[\mu_H^j > \mu_H^i] + \tfrac{1}{2} w_H \cdot 1[\mu_H^i = \mu_H^j]
\]

The \textbf{confidence score} for the comparison is:
\[
\text{conf}(M_i, M_j) = \left| V(M_i, M_j) - V(M_j, M_i) \right| \in [0,1],
\]
which quantifies the net strength of weighted human preference between the models.

To obtain an acyclic partial order within the cluster, we:
\begin{enumerate}
    \item Form a set of directed edges $(M_i \to M_j)$ for all pairs with $V(M_i, M_j) > V(M_j, M_i)$, assigning each edge strength $s(i,j) = \text{conf}(M_i, M_j)$.
    \item Sort edges by descending $s(i,j)$, breaking ties deterministically (e.g., by $|\mu_i - \mu_j|$ or model index).
    \item Initialize a directed graph $G$ over the models in the cluster, with no edges.
    \item Process edges in sorted order, adding each to $G$ only if it does not create a directed cycle.
\end{enumerate}

The resulting $G$ is a directed acyclic graph: if a path exists from $M_i$ to $M_j$, we conclude $M_i \succ M_j$; if no path exists in either direction, the models remain tied.

\medskip
\noindent
This design uses the full information from weighted human judgments, prioritizes the strongest and most reliable preferences, and ensures the absence of cycles. It refines ties when the evidence is clear, but preserves ties when preferences are inconclusive or contradictory, preventing overinterpretation of sparse evaluation data.

\subsection{Align-Score Metric}
We define an alignment score $\mathit{align\text{-}score} \in [0,1]$ between a human ranking $R_H$ and a judge ranking $R_J$. It is based on an \emph{evaluation discrepancy} $\epsilon_{\text{rank}}$ that combines two components: a confidence-weighted rank disagreement capturing how often and how strongly $R_J$ differs from $R_H$, and a normalized score error measuring differences in assigned quality scores. The alignment score is computed as  $\mathit{align\text{-}score} = 1 - \epsilon_{\text{rank}}$,
so higher values indicate stronger agreement.

\noindent
The {\it confidence-weighted rank disagreement} is an adaptation of Kendall’s $\tau$ distance, where each pairwise inversion is weighted by the human-assigned confidence for that model pair.
% For each  pair of models $M_i, M_j$ from the ordering in $R_H$ and $R_J$:
% -- If $R_H$ prefers $M_i \succ M_j$ and $R_J$ violates this, penalize by $\text{conf}(M_i, M_j)$.\\
% -- If $R_H$ prefers $M_i \succ M_j$ and in $R_J$ the order is undecided, penalize by $\text{conf}(M_i, M_j)$.\\
% -- If in $R_H$ the order between $M_i$ and $M_j$ is undecided (confidence level is 0) but $R_J$ prefers one model over the other, penalize by $1$.\\
% -- If both rankings are indifferent, apply no penalty.
% The total penalty is normalized by the number of model pairs.

For each pair of models $(M_i, M_j)$ in the ordering set \\
$p_{ij}=\text{conf}(M_i, M_j)$ if $R_H$ prefers $M_i \succ M_j$ and $R_J$ reverses or ties; \\
$p_{ij}=1$ if $\text{conf}(M_i, M_j)=0$ and $R_J$ imposes an order; 
otherwise $p_{ij}=0$. \\
The rank disagreement is 
$\epsilon_{\text{rank}}=\frac{\sum_{i<j} p_{ij}}{\binom{k}{2}}$,
where $k$ is the number of models.

For the score error, let $s_H(M_i)$ and $s_J(M_i)$ denote the mean scores assigned to model $M_i$ by humans and by the judge respectively, both normalized to $[0,1]$.  
The normalized score error is defined as:
\[
\epsilon_{\text{score}} =
\frac{1}{k} \sum_{i=1}^{k} \left| s_H(M_i) - s_J(M_i) \right|.
\]

The overall total evaluation discrepancy is:
\[
\varepsilon = \alpha \cdot \epsilon_{\text{rank}} + (1 - \alpha) \cdot \epsilon_{\text{score}},
\]
where $\alpha \in [0,1]$ controls the trade-off between ranking preservation and score proximity.

\subsection{Experimental Results: COBOL Code Explanation
}
We applied the SparseAlign framework to the COBOL code explanation task to evaluate how well two candidate judges aligned with expert human assessments under extremely limited annotation. 

The dataset contained COBOL code snippets extracted from production applications, along with natural-language explanations generated by LLM models under active development by model-training teams.
These evaluations were conducted to inform decisions on which models demonstrated superior performance and should be advanced toward production deployment. Each explanation was reviewed by our subject-matter experts (SMEs) according to task-specific evaluation guidelines. 
The evaluation used a 1–7 scoring scale.  A total of 7 human evaluators ($H_1, \ldots, H_7)$ assessed explanations produced by 6 different models ($M_1, \ldots M_6)$. Each model generated explanations for the same 25 code snippets, resulting in 150 (code, explanation) pairs. These pairs were distributed across annotators without overlap, as illustrated in Figure~\ref{fig:human_eval_data}.

For this experiment, we evaluated four candidate LaaJs. 
$L_1$, based on \texttt{llama-405b-instruct}, and  
$L_2$, based on \texttt{llama-4-maverick}, were both configured with the same evaluation prompt (omitted here for brevity) to isolate the effect of the underlying LLM’s capabilities while keeping evaluation criteria fixed.

We also included two simulated baselines. 
$L_{\mathrm{random}}$ assigns scores uniformly at random in the range 1–7, representing a completely uninformative judge.  
$L_{\mathrm{human\_close}}$ is derived from the human evaluation table by perturbing 10\% of the grades in each model column by either $+1$ or $-1$ (clipped to $[1,7]$). This retains the overall human preference structure while introducing controlled noise, simulating a high-quality but imperfect judge. These simulated LaaJs serve as stress tests for our framework: $L_{\mathrm{random}}$ provides a clear negative control and receives a low alignment score, while $L_{\mathrm{human\_close}}$ achieves high alignment as expected. This validates SparseAlign’s ability to reliably distinguish between strong and weak evaluators.

First, we derived the human reference ranking using the \textit{Human-Rank} algorithm. Table~\ref{tab:human_scores} reports the mean and standard deviation of human evaluation scores for each model.

\begin{table}[h]
\centering
\caption{Mean and standard deviation of human scores per model.}
\label{tab:human_scores}
\begin{tabular}{lcccccc}
\toprule
 & M1 & M2 & M3 & M4 & M5 & M6 \\
\midrule
\textbf{Average} & 5.08 & 4.80 & 4.857 & 4.20 & 4.52 & 5.08 \\
\textbf{Std. Dev.} & 1.129 & 1.470 & 1.457 & 1.720 & 0.900 & 1.324 \\
\bottomrule
\end{tabular}
\end{table}

We continue with computing the threshold 
$\delta = \frac{1}{6} \cdot \text{median}(\sigma_i) \sim 0.23$. This threshold determines when differences in mean scores are too small to indicate a meaningful preference.
Applying it to the mean human scores produces the following ranking, from most to least preferred, with ties: 
$< \{M_1, M_6\}, \{M_3, M_2\}, M_5, M_4 >$.

\begin{table}[h]
\centering
\caption{Average Scores of Models M1 and M6 per Human Evaluator}
\begin{tabular}{lcc}
\hline
\textbf{Evaluator} & \textbf{Av. Score M1} & \textbf{Av. Score M6} \\
\hline
H1: 4/25  & \textbf{4.75} & 4.25 \\
H2: 1/25  & 5.00 & \textbf{6.00} \\
H3: 2/25  & \textbf{6.00} & 5.50 \\
H4: 3/25  & 4.00 & \textbf{5.30} \\
H5: 3/25  & \textbf{6.00} & 4.00 \\
H7: 10/25 & 5.20 & \textbf{5.70} \\
H8: 2/25  & \textbf{4.50} & 4.00 \\
\hline
\end{tabular}
\end{table}

We then apply the weighted voting scheme to resolve the ties between $M_1$ and $M_6$, and between $M_3$ and $M_2$. For $\{M_1, M_6\}$, the total normalized votes are $V(M_1) = 4/25 + 2/25 + 3/25 + 2/25 = 0.44$ and $V(M_6) = 1/25 + 3/25 + 10/25 = 0.56$, yielding $M_6 \succ M_1$ with confidence $0.12$. For $\{M_3, M_2\}$, we obtain $V(M_2) = 0.52$ and $V(M_3) = 0.48$, giving $M_2 \succ M_3$ with confidence $0.04$.

Overall, the \textit{Human-Rank} algorithm yields the following human reference ranking:
$
R_H = \langle M_6,\, M_1,\, M_2,\, M_3,\, M_5,\, M_4 \rangle
$,
with $\text{conf}(M_6, M_1) = 0.12$ and $\text{conf}(M_2, M_3) = 0.04$, while all other model pairs have $\text{conf} = 1$.

Next, we ran all the LaaJs on the same set of samples, computed the average scores per model, and derived the corresponding rankings (omitted from submission).

\begin{figure}[h]
    \centering
    \includegraphics[width=0.9\linewidth]{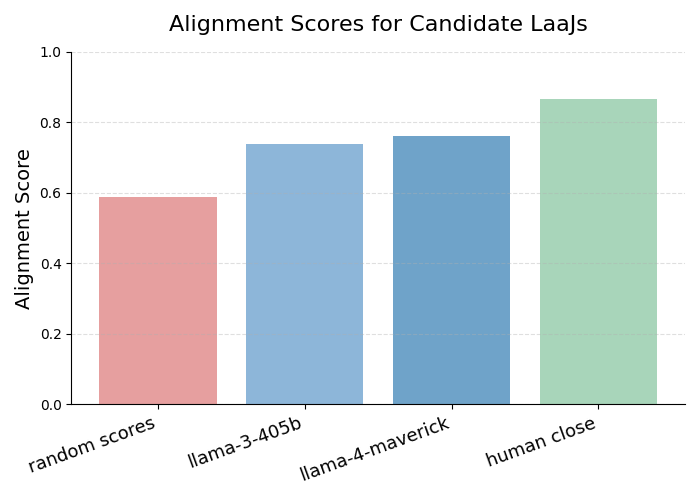}
    \caption{SparseAlign correctly assigns low alignment to the weak evaluator $L_{\mathrm{random}}$ 
and high alignment to the strong evaluator $L_{\mathrm{human\_close}}$, 
while quantifying the performance of the two real LaaJs.}
    \label{fig:align_scores}
\end{figure}

% \[
% R_{L_1} = \langle \{M_6, M_2\}, M_3, M_5, \{ M_4, M_1\} \rangle.
% \]
% \[
% R_{L_2} = \langle \{ M_6, M_2, M_3\}, M_1, M_5, M_4 \rangle.
% \]
% \[
% R_{L_{\mathrm{random}}} = \langle \{ M_1, M_4, M_6 \}, \{ M_2, M_5 \}, M_3 \rangle.
% \]
% \[
% R_{L_{\mathrm{human\_close}}} = \langle \{ M_1, M_6, M_3 \}, \{ M_2, M_5 \}, M_4 \rangle.
% \]
Applying the align-score metric with the human reference ranking $R_H$, we decomposed the evaluation discrepancy $\epsilon$ into its two components: 
confidence-weighted rank disagreement $\epsilon_{\text{rank}}$ and normalized score error $\epsilon_{\text{score}}$. 
The results for each candidate LaaJ are shown in Table~\ref{tab:align_results} and Figure \ref{fig:align_scores}.

\begin{table}[h]
\centering
% \caption{Alignment score decomposition for the candidate LaaJs}
\caption{}
\vspace{0.5em}

\begin{tabular}{lccc}
\toprule
\textbf{LaaJ} & $\epsilon_{\text{rank}}$ & $\epsilon_{\text{score}}$ & \textbf{$\mathit{align\text{-}score}$} \\
\midrule
Llama-4-maverick   & 0.30  & 0.18  & \textbf{0.76} \\
Llama-3-405b        & 0.38  & 0.14  & \textbf{0.74} \\
Random Scores      & 0.475 & 0.347 & \textbf{0.589} \\
Human\_Close       & 0.211 & 0.058 & \textbf{0.866} \\
\bottomrule
\end{tabular}
\label{tab:align_results}
\end{table}

The results show that both candidate LaaJs achieve high alignment with human judgments, with Llama-4-maverick slightly outperforming Llama-3-405b due to lower rank disagreement. The human-close LaaJ scores highest, as expected, while the random LaaJ scores substantially lower, confirming SparseAlign’s ability to distinguish between high- and low-fidelity evaluators.

\bibliographystyle{plain}  % or other styles: ieeetr, abbrv, unsrt, etc.
\bibliography{references} 

\end{document}